\documentclass[11pt]{article}
\usepackage{color}
\usepackage[toc]{appendix}
\usepackage{amsmath}
\usepackage{url}

\begin{document}
\setcounter{page}{1}
\def\theequation{\arabic{section}.\arabic{equation}} 
\newcommand{\be}{\begin{equation}}
\newcommand{\ee}{\end{equation}}
\newcommand{\ul}{\underline}
\hyphenation{Her-nan-dez}
\begin{titlepage}
\title{Foliation dependence of black hole 
apparent horizons \\ in spherical symmetry}
\author{Valerio 
Faraoni,$^a$\footnote{vfaraoni@ubishops.ca} \space George F.R. 
Ellis,$^b$\footnote{gfrellisf@gmail.com} \space Javad T. 
Firouzjaee,$^{c,d}$\footnote{j.taghizadeh.f@ipm.ir} \\ 
Alexis Helou,$^{e}$\footnote{alexis.helou@physik.uni-muenchen.de}  
\space and 
Ilia Musco$^f$\footnote{ilia.musco@obspm.fr}\vspace{0.05cm} \\
{\small $^a$ Physics 
Department, Bishop's University}\\
{\small 2600 College Street, Sherbrooke, Qu\'ebec, Canada 
J1M~1Z7}\\
{\small $^b$~Mathematics and Applied Mathematics 
Department}\\ 
{\small University of Cape Town, Rondebosch, Cape Town 
7701, South Africa}\\
{\small $^c$ School of 
Astronomy, Institute for 
Research in Fundamental Sciences (IPM)}\\
{\small P.O. Box 19395-5531, Tehran, Iran}\\
{\small $^d$ Department of Physics (Astrophysics), 
University of Oxford}
{\small Keble Road, Oxford OX1 3RH, UK}\\
{\small $^e$ Arnold Sommerfeld Center, 
Ludwig-Maximilians-Universit\"at,}
{\small  Theresienstr. 37, 80333 M\"unchen, Germany}\\
{\small $^f$Laboratoire Univers et Th\'{e}ories, UMR 8102 CNRS, Observatoire de 
Paris, Universit\'{e} Paris Diderot}\\
{\small 5 Place Jules Janssen, F-92190 Meudon, France}
}

\date{} 
\maketitle
\thispagestyle{empty}
\begin{abstract}

Numerical studies of gravitational collapse to black holes 
make use of apparent horizons, which are intrinsically 
foliation-dependent. We expose the problem and discuss 
possible solutions using the Hawking quasilocal mass. In spherical symmetry, 
we present a physically sensible approach to the problem by restricting to 
spherically symmetric spacetime slicings. In spherical symmetry the 
apparent horizons are gauge-independent in any spherically symmetric
foliation but physical 
quantities associated with them, such as surface gravity and 
temperature, are not. The widely used  
comoving and Kodama foliations, which are of particular interest, 
 are discussed in detail.

\end{abstract}  \end{titlepage}   
\clearpage

\section{Introduction}
\label{sec:1}
\setcounter{equation}{0}

Realistic black holes interact with their environment and 
are, therefore, 
dynamical. The gravitational collapse 
leading to black hole formation is also a highly dynamical 
process. In time-dependent situations the event horizons 
(which are null surfaces) familiar from the study of 
stationary black holes \cite{Wald, FrolovNovikov, Poisson} 
are replaced in practice by apparent horizons, which can have timelike, lightlike 
or spacelike nature ({\em e.g.}, \cite{Boothreview, Booth:2005ng,
Nielsenreview, mybook}).
These are defined as the locus of vanishing expansion of a null geodesic congruence emanating from a spacelike compact 2-surface ${\cal S}$ with spherical topology.\\

In contrast with the event horizon, which is a global concept 
defined using the global structure of spacetime, the apparent horizon is 
a {\em quasilocal} concept.  In numerical studies of collapse,
it is more practical to track  
apparent horizons, rather than event horizons which require the knowledge
of the entire future history of the spacetime\footnote{See, 
however, Refs.~\cite{Diener2003, CohenPfeifferScheel, 
Thornburg, Lousto}.} \cite{Thornburg, BaumgarteShapiro, 
Chuetal}. A significant fraction of research in 
numerical relativity aims at predicting with high precision the 
waveforms of gravitational waves generated in 
the merger of compact-object binary systems or in stellar 
collapse to form black holes. These waveforms enter 
data banks for use in the laser interferometric detection 
of gravitational waves. Comparison with templates played a 
crucial role in the recent observations of gravitational 
waves from 
black hole mergers by the {\em LIGO} collaboration 
\cite{LIGO, LIGO2}. These numerical works also 
use apparent horizons.\\

Apparent horizons suffer from a drawback: in general, they 
are foliation-dependent, because they depend on the choice 
of the 2-surface ${\cal S}$, which is chosen to lie in 
some hypersurface ${\cal H}$ that is a surface of 
simultaneity for some family of observers ${\cal O}$. This 
is a problem since the existence of a  
horizon is ultimately  the defining feature of 
a black hole\footnote{In Rindler's words, a horizon is ``a 
frontier between things observable and things 
unobservable'' \cite{Rindler}.}
and this means that, in 
dynamical situations, the  defining feature of a black hole used in practice 
depends on the observer.  {\em A priori} the situation 
seems actually worse: even the {\em existence} of a black 
hole seems to depend on the observer, as epitomized by the 
fact that the Schwarzschild spacetime (the prototypical 
black hole geometry) has no apparent horizons in certain 
foliations \cite{WaldIyer, SchnetterKrishnan}, giving the 
impression that cosmic censorship is violated.  The basic 
idea of Ref.~\cite{WaldIyer} is this: the Schwarzschild-Kruskal 
geometry admits ``angular horizons''. For example, the 
North pole of a 2-sphere inside the Schwarzschild black 
hole cannot send light signals to events with angular 
coordinate $\theta$ larger than a critical value, for 
example at the South pole of a 2-sphere. These North and South 
poles are spacelike-related. Hence, it is possible (as 
shown in \cite{WaldIyer}, see also \cite{SchnetterKrishnan}) 
to construct spacelike Cauchy 
surfaces which interpolate between trajectories of the 
North pole, which come close to the $r=0$ singularity, and 
trajectories of the South pole, which remain outside the black 
hole. The causal past of any Cauchy surface in a slicing so 
constructed contains no trapped surfaces, yet it comes 
arbitrarily close to the singularity. Needless to say, 
these slicings are highly non-spherically symmetric and certainly 
contrived, but the problem of principle remains that 
foliations containing no apparent horizons exist even in 
the Schwarzschild geometry.\\

Is the situation really that bad? The problem would be 
ameliorated, if not solved, if there were preferred 
foliations of spacetime, that is, if the spacetime slicing 
was somehow fixed  by physical arguments such as symmetries. Fixing the spacetime foliation is 
ultimately equivalent to fixing a family of observers, and 
this is already a practical necessity in certain problems 
related with black hole physics. For example, the 
temperature of a black hole calculated with quantum 
field theory (for a scalar field) in curved space depends 
on the vacuum state, which in turn 
depends on which observer is chosen since the particle 
number operator is not invariant under change of 
frame. \\

How do we fix the spacetime slicing in a physically 
meaningful way? We do not consider here Lorentz-violating 
theories which have a preferred frame, restricting 
ourselves to general relativity. In the presence of 
spacetime symmetry, a foliation which respects this symmetry is 
certainly preferred from both the geometric and physical 
points of view. For example, in 
Friedmann-Lema\^itre-Robertson-Walker universes, preferred 
observers are those associated with spatial homogeneity and 
isotropy, {\em i.e.}, those who see the cosmic microwave 
background as being homogeneous and isotropic around them (apart 
from the well known tiny temperature fluctuations with 
$\delta T/T \sim 10^{-5}$). Similarly, we argue that in the 
presence of spherical symmetry,
foliations 
which preserve spherical symmetry are naturally preferred 
and that, with this restriction, the gauge-dependence 
problem of apparent horizons is circumvented, but the 
thermodynamics of these horizons remains 
gauge-dependent. The reason for 
this gauge-independence can be summarized in the fact that, 
in spherical symmetry, the areal radius $R$ is a 
geometrically well-defined quantity and the apparent 
horizons are located by the scalar equation\footnote{We 
refer to black hole apparent 
horizons but most of our considerations apply to 
cosmological apparent horizons as well.}
\begin{equation}
\nabla_c R \, \nabla^c R=0.
\end{equation}
 The gauge-independence of apparent horizons 
is verified for the Kodama and comoving gauges, which are 
probably the ones most used in the 
literature.
The argument applies also to the computation of the 
Misner-Sharp-Hernandez quasilocal mass 
$M_\text{MSH}$ which is widely used in spherical 
collapse and in the thermodynamics of apparent 
horizons.\footnote{The Misner-Sharp-Hernandez mass 
is also used for cosmological black holes  
\cite{Firouzjaee:2010ia, mybook} and coincides with the 
Lema\^itre mass used in Lema\^itre-Tolman-Bondi geometries 
\cite{Lemaitremass}.} 
When evaluated on an apparent horizon, this quantity 
satisfies the scalar equation  \cite{MSH2}
\begin{equation}
 R_\text{AH}=2M_\text{MSH}\left( R_\text{AH}\right).
\end{equation}

More generally, studying black holes from the dynamical 
point of view requires specifying the black hole mass, one 
of its fundamental parameters. In the different context of 
black hole (or horizon) thermodynamics, the black hole mass 
plays the role of internal energy in the first law of 
thermodynamics. Most of the literature on black holes 
assumes that this mass is defined by the Hawking quasilocal 
mass \cite{Hawking} computed at the apparent horizon. 
In spherical symmetry, the Hawking 
 quasilocal mass reduces to the 
 Misner-Sharp-Hernandez mass \cite{MSH, MSH2} 
used in studies 
of spherical collapse.\\

Our goal in this article is not to prove new theorems with 
full rigor: instead, we adopt a more pragmatic approach 
(in the philosophy of \cite{Visser}) 
to the issue of apparent horizons and look into 
possible ways of 
alleviating the problems. While rigorous mathematical 
results (see \cite{Hayward:1993wb, Hayward:1997, Eardley:1997hk, BenDov:2006vw, Ashtekar:2005ez, Booth:2005qc, Bengtsson:2010tj} for already established examples) 
are ultimately desirable, a more pragmatic approach 
may be convenient to move forward in practical 
applications, leaving the problems of 
principle to be attacked in the future. In the next section 
we briefly review the dual role that the Hawking 
quasilocal mass plays as the internal energy in the first law of black 
hole thermodynamics and as the physical mass in dynamical black hole 
spacetimes. We discuss how this quasilocal construct could select a 
foliation, but this choice is useless for practical purposes. To 
proceed, we restrict to spherical symmetry in Section~\ref{sec:3} and we 
discuss two foliations, the comoving (Landau) and the Kodama gauges, and 
the explicit relation between them. We show how the apparent horizons 
coincide, as geometric surfaces, but have different thermodynamic interpretations 
according to the comoving and Kodama observers.
We use metric signature $-+++$ and units in which the speed 
of light and Newton's constant are unity (and, when 
discussing thermodynamics, we use 
geometrized units in which also the reduced Planck constant 
and the Boltzmann constant are unity), and we 
otherwise follow the notation of Wald's book \cite{Wald}.

\section{Hawking mass and apparent horizons}
\label{sec:2}
\setcounter{equation}{0}

\subsection{Hawking mass and a preferred foliation}

In a spacetime $\left( {\cal M}, g_{ab} \right)$ which is a 
solution of the Einstein equations, let ${\cal S}$ be a 
2-dimensional, spacelike, embedded, compact, and orientable 
surface  
which is chosen to lie in some hypersurface ${\cal H}$ 
that is a surface of simultaneity for some family of 
observers ${\cal O}$. The 4-dimensional spacetime metric 
$g_{ab}$ induces a 2-metric $h_{ab}$ on ${\cal S}$, with 
${\cal R}^{(h)}$ being the Ricci scalar of $h_{ab}$. One 
can consider the congruences of 
ingoing ($-$) and outgoing ($+$) null geodesics $k^a_\pm$ 
at the 
surface ${\cal S}$. Denote with $\theta_{(\pm)}$ and 
$\sigma_{ab}^{(\pm)}$ the expansion scalars and the shear 
tensors of these null geodesic congruences, respectively. 
The Hawking quasilocal mass associated with the  
2-surface ${\cal S}$ is the integral quantity  
\cite{Hawking}. 

\begin{equation}
M_\text{H} = \frac{1}{8\pi G} \sqrt{ \frac{A}{16\pi}} 
\int_{\cal S} 
\mu \left( {\cal R}^{(h)} +\theta_{(+)} \theta_{(-)} 
-\frac{1}{2} \, \sigma_{ab}^{(+)} 
\sigma^{ab}_{(-)}    
\right) \,, 
\label{HHmass}
\end{equation}
where $\mu$ is the volume 2-form on ${\cal S}$ and $A$ is 
the area of ${\cal S}$.  
Consistent with spherical symmetry, we will assume that the 
surface ${\cal S}$ is a topological 2-sphere. 

Clearly, the choice of the surface ${\cal S}$ is essential 
in both the construction~(\ref{HHmass}) and in the 
computation of $M_\text{H}$. The value of this quantity 
depends on the choice of the unit normal $n^a$ to ${\cal 
S}$. 

\subsection{Fixing the foliation: method~1 (observer 
four-velocity)}

One possible way to fix the foliation is to  
identify the (timelike) 4-velocity $u^a$ of an 
``observer'' at ${\cal S}$ with $n^a$ (remember that 
${\cal S}$ is spacelike). A foliation of 
spacetime is 
associated with a family of observers. A possible (maybe 
even natural)  choice 
for this observer is picking the one that ``sees'' 
the hypersurface ${\cal H}$ containing ${\cal S}$ and the matter on it at rest, implying ${\cal S}$ will also be at rest. 
If the unit normal 
$n^a$ to the surface ${\cal S}$ has only its time 
component different from zero, then the surface ${\cal S}$ 
is at rest in these coordinates. One can then foliate the 
3-dimensional  space ${\cal H}$ by   
carrying points of the surface 
${\cal S}$ along the normal as done in the construction 
of Gaussian normal 
coordinates ({\em e.g.}, \cite{Wald}). There will be some kind of coordinate singularity as one approaches the centre of ${\cal S}$.\\

The Hawking mass $M_\text{H}$ should be the time component of a timelike 
4-vector $P^a$ describing the energy and 3-momentum of a 
certain 
region of spacetime enclosed by the compact surface ${\cal 
S}$ \cite{Szabados}. Therefore, the zero component $P^0$ 
of this vector should be  gauge-dependent.  As, in special 
and general relativity, the ``mass of a particle'' is 
identified with 
its rest mass, the zero component of its 4-momentum in the 
frame in 
which the particle is at rest, it would be natural to 
identify the Hawking quasilocal mass associated with a 
compact spacelike 2-surface ${\cal 
S}$ with the energy~(\ref{HHmass}) seen by an observer for which 
${\cal S}$ is at rest. A different observer in 
motion with respect to ${\cal S}$ would ascribe a different 
value to the mass of this surface. If it were accepted that 
the Hawking mass describes the mass-energy seen by an 
observer at rest with respect to the surface ${\cal S}$, 
then there would be no room for considering spacetime 
foliations defined by different observers $u^a \neq n^a$.\\ 
 
Can this be the ``right'' procedure to fix the foliation? 
The definition of Hawking mass~(\ref{HHmass})  
applies to {\em spacelike} (compact) 2-surfaces.
In practice, in black hole studies, the surface ${\cal S}$ 
is chosen to be the two-dimensional intersection of 
the black hole apparent horizon (a three-dimensional 
world tube, generated by a vector field $t^a$) 
with a time slice (a spacelike 3-surface ${\cal 
H}$).\footnote{Note that, in the literature, 
the terminology ``apparent 
horizon'' may refer to both the three-dimensional world tube and its 
two-dimensional cross-section.}
 Three-dimensional apparent horizons can be spacelike, null or timelike 
but their intersections with hypersurfaces of 
constant time ${\cal H}$ are two-dimensional spacelike surfaces, hence the definition applies. 
There are two independent null normal vectors $k^a_\pm$ to the surface ${\cal S}$. There is also a  
spacelike normal $s^a$ lying in the time slice ${\cal H}$ containing ${\cal S}$, and a timelike  
normal vector $n^a$ to ${\cal S}$ that is normal to ${\cal H}$. 
This works even when the apparent horizon 
is null or spacelike, as it does not depend on its causal nature.  
The  identification $u^a=n^a$ as the normal to ${\cal S}$ leads to the 
interpretation of $M_\text{H}$ as ``the mass seen by this observer''. 
But, if we also demand that our observer be at rest with respect 
to the apparent horizon ${\cal S}$, then we must have $u^a\propto t^a$, 
with $t^a$ the tangent to the three-dimensional horizon world-tube.
The latter can be null or spacelike, in which case 
the above interpretation of $M_H$ becomes moot.
Moreover, in practical calculations of collapse the apparent horizon is not at 
rest in the reference frame of interest.\footnote{This is 
the case, for example, when comoving coordinates are used 
while the apparent horizon is not comoving.} 
We conclude that this  method does not seem suitable. 
\\

However it must make sense also for other observers, who do 
not see ${\cal S}$ at rest, to speak of the mass of a black 
hole. As an alternative, it is possible to consider a 
foliation with $u^a\neq n^a$ specified in some different 
way, as we do in the next section for the special case of 
spherical symmetry.
Once a foliation is fixed, one cannot use another 
slicing in the same calculation. In general spacetimes 
one does not know how to fix the foliation, but this 
is possible in spherical symmetry, to which we devote 
the rest of this work.

\section{Spherical symmetry}
\label{sec:3}
\setcounter{equation}{0}
 The assumption of spherical symmetry facilitates the 
analytical study of black holes; it allows modelling  
them with a set of appropriate tools such as the 
areal radius, the Misner-Sharp-Hernandez mass, 
the Kodama vector field,  and the Hayward-Kodama 
surface gravity.
\\

Therefore let us  specialize the previous discussion to spherical 
symmetry. In a spherically symmetric spacetime, the 
Hawking mass reduces to the Misner-Sharp-Hernandez mass 
$M_\text{MSH}$ \cite{Haywardspherical}.  
If $R$ is the areal radius of the spherical geometry, then 
$M_\text{MSH}$ is defined by \cite{MSH2, Haywardspherical}
\be\label{MSH}
1-\frac{2M_\text{MSH}(R)}{R}=\nabla_cR \, \nabla^cR 
\ee
and the apparent horizons (when they exist) are the roots 
of the equation
\be \label{locusAH}
\nabla_cR \, \nabla^cR=0 \,,
\ee
so that 
\be \label{R=2M}
R_\text{AH}=2M_\text{MSH}(R_\text{AH})
\ee
at the apparent horizons. This relation generalizes the one 
holding for mass and radius of the Schwarzschild event 
horizon.\\

Before proceeding, one should note that the areal radius 
$R$ is a geometric quantity \cite{MSH2} which is independent of the 
foliation $\{\cal H\}$ which determines $\{\cal S\}$, 
and eq.~(\ref{locusAH}) locating the apparent 
horizons is a {\em scalar} equation.
Therefore, all {\em spherically symmetric} foliations 
will produce the same apparent horizons 
(but this is not true of non-spherical 
foliations, as demonstrated by the example of 
\cite{WaldIyer}). In other words, these horizons are 
gauge-independent if one restricts to spherically symmetric slicings. 
Similarly, the Hawking mass is gauge-dependent 
in general, but in spherical symmetry 
it becomes gauge-independent because it 
is defined by the scalar equation~(\ref{MSH}).

In spherical symmetry one is usually interested in 
computing the Hawking/Misner-Sharp-Hernandez quasilocal mass when the surface 
${\cal S}$ coincides with a 2-sphere of symmetry with unit 
normal $n^{a}$. Usually one needs to compute the Hawking 
mass {\em on the apparent horizon}, which depends on the 
foliation chosen. 
A crucial point of our discussion is that it is natural 
to consider spacetime foliations which are spherically 
symmetric and it seems contrived to do otherwise and 
not take into account the undeniable simplifications 
brought about by the symmetries of the spacetime 
under study.
Then, a foliation will be identified with a family of observers 
$u^a$ which have only time and radial components, 
$u^{a}=\left( u^0, u^1, 0,0 \right)$ in 
coordinates adapted to the symmetry. The 2-sphere ${\cal 
S}$ can have radial motion in this foliation, {\em i.e.}, 
$n^{a}=\left( n^0, n^1, 0,0 \right)$ with $n_c n^c=-1$ in 
these coordinates.\footnote{In particular, if the surface 
${\cal S}$ is at rest in this foliation, $n^{a}=\left( 
1/\sqrt{|g_{00}|},0,0,0 \right)$ in adapted coordinates.}\\

In the following we discuss explicitly  two families of observers 
which have the prescribed spherically symmetric form $u^{a}=\left( u^0, u^1, 0,0 \right)$:
the comoving (Landau) gauge, and the Kodama foliation.

\subsection{Fixing the foliation: method~2 (comoving gauge)}

Numerical studies of the spherical collapse of a fluid to a 
black hole have often employed the comoving gauge, that is, the 
timelike 4-vector $u^a$ describing the foliation is 
identified with the 4-velocity of the collapsing 
fluid.\footnote{In general, the latter is not parallel to 
the Kodama vector.} This procedure fixes the spacetime 
foliation in a way alternative to the previous one.\\

The energy momentum tensor of a perfect fluid\footnote{We recall 
that a perfect fluid is a fluid with no dissipative effects 
(no shear and no viscosity).} has the form
\be
T^{ab}=\left( \rho+p \right) u^a u^b+ p g^{ab} \,,
\label{perfect_fluid}
\ee
where the fluid 4-velocity $\left( u^t, u^t v^i \right)$ is 
normalized as $u_c u^c=-1$ and $v^i$ is its 3-velocity. 
Misner and Sharp \cite{MSH} write their equation in 
spherically symmetric form setting   the shift $N^i =0 $ 
and with 3-velocity $v^i=0$. In other words, they fix the foliation by 
assuming that the normal to the hypersurfaces $t=\text{const}$.\
coincides with the fluid 4-velocity (comoving 
gauge).\footnote{This is possible because spherical 
symmetry implies the vorticity is zero.}  
The line element in comoving gauge is
\be\label{comovinggauge}
ds^2= -\mbox{e}^{2\phi} dt^2 +\mbox{e}^{\lambda} dr^2 +R^2 d\Omega_{(2)}^2 \,,
\ee
where $\phi$, $\lambda$ and $R$ (the areal radius) are 
all functions of $t$ and $r$, and 
$d\Omega_{(2)}^2=d\theta^2 +\sin^2 \theta \, d\varphi^2$ is 
the line element on the unit 2-sphere. 
 (See Appendix \ref{appendix} for examples of known spacetimes 
expressed in the comoving gauge.) It can be shown that 
there is a special slicing $\left(t',r' \right)$ such that 
the energy flux vanishes relative to the fluid 4-velocity, thus 
\begin{equation}
u_a T^{ab} \left( g_{bc}+u_b u_c \right)=0 
\end{equation} holds  in this 
frame (called the Landau or energy frame). Here $g_{bc}+u_b 
u_c \equiv \gamma_{bc}$ is the Riemannian metric on the 
3-space 
and, by the Einstein field equations, $u^a$ is then a Ricci 
eigenvector so the Landau gauge is uniquely defined 
geometrically. 
 (Choosing the Kodama 
gauge, as discussed in Section \ref{section_Kodama_gauge}, 
does not give zero heat flux for the matter fluid, which 
shows that the Kodama gauge is different from the 
energy frame.)

Consider a general 2-sphere of symmetry ${\cal S}$, for 
which one writes the associated Misner-Sharp-Hernandez 
mass. This surface is either at rest or 
expanding/contracting radially with respect to the fluid.  
Choosing the comoving gauge eliminates by {\em 
fiat} the question of foliation-dependence of the 
apparent horizons. 
Apparent horizons (which are 2-spheres once we 
restrict to spherically symmetric foliations) 
are almost never comoving, and thus move 
radially with respect to observers 
comoving with the fluid. The comoving observer with 
4-velocity  $u^c$ does not coincide with the unit normal 
$n^c$ to the surface ${\cal S}$ at the apparent horizon.\\

 Let us now introduce some tools adapted to the comoving 
observer, 
in order to make contact with the literature 
(we will use them again in Section~\ref{sec:4}).
Following Ref.~\cite{HelouMuscoMiller2016}, we define the 
derivatives with respect to proper 
time and proper radial distance as measured by the comoving 
observers 
\be
D_t \equiv \mbox{e}^{-\phi} \partial_{t} \,, \;\;\;\;\;\;
D_r \equiv \mbox{e}^{-\lambda/2} \partial_{r} \,.
\ee
Applying these derivative operators to the areal radius 
one obtains
\begin{eqnarray}
U &\equiv & D_t R=\mbox{e}^{-\phi} \, \frac{\partial 
R}{\partial t}  \equiv 
\mbox{e}^{-\phi} \dot{R} \,,\label{U}\\
&&\nonumber\\
\Gamma &\equiv & D_r R=\mbox{e}^{-\lambda/2} 
\, \frac{\partial R}{\partial r}  
\equiv  \mbox{e}^{-\lambda/2} R' \,,\label{Gamma}
\end{eqnarray}
The apparent horizons of the geometry~(\ref{comovinggauge})  
are located by the roots of eq.~(\ref{locusAH}), 
which yields \cite{HelouMuscoMiller2016}
\be\label{Usquare}
\Gamma^2 -U^2 =0
\ee
with $U=-\Gamma$ corresponding to black hole apparent horizons 
and $U=\Gamma$ to cosmological apparent horizons 
 (for an expanding universe), respectively. 
Using the quantities $U$ and $\Gamma$, the line 
element~(\ref{comovinggauge}) is rewritten as 
\be
ds^2=-\frac{\dot{R}^2}{U^2} \, dt^2 +\frac{R'^2}{\Gamma^2} 
\, dr^2 +R^2 d\Omega_{(2)}^2
\ee
and the equation $\nabla_cR \, \nabla^c R=0$ locating the 
apparent horizons, of course, reproduces 
eq.~(\ref{Usquare}). 

The three-dimensional velocity $v$ of an observer with 
respect to the matter is obtained from the derivative of the proper 
radial distance with respect to the proper time of comoving 
observers, given by \cite{HelouMuscoMiller2016}
\be\label{threevelocity}
\frac{dR}{d\tau}=\mbox{e}^{-\phi} \left( \dot{R}+R' 
\, \frac{dr}{dt}\right) 
=U+\Gamma v \,.
\ee
 In particular, one may evaluate the above at the horizon 
to get $v_H^{(C)}$, the three-velocity of the horizon 
with respect to the comoving observer. 
Using Einstein equations with \eqref{perfect_fluid}, 
it yields \cite{HelouMuscoMiller2016}
\begin{equation}
 v_H^{(C)} = \frac{1+8\pi R_{AH}^2p}{1-8\pi R_{AH}^2\rho}  \ ,
 \label{eq_vH_BH}
\end{equation}
where the energy density and pressure are evaluated at the 
horizon location.
 
  Finally, the four-velocity of comoving 
observers has components 
\be\label{ucomoving}
u^{\mu}_{(C)}=\left( \mbox{e}^{-\phi}, 0,0,0 \right)
\ee
in comoving coordinates $\left( t,r,\theta, \varphi 
\right)$.  This will be compared to the four-velocity 
of the Kodama observer -- as introduced below --
in Section~\ref{sec:4}.

\subsection{Fixing the foliation: method~3 (Kodama gauge)}
 \label{section_Kodama_gauge}

\subsubsection{Kodama vector field and Kodama gauge} 

 In general relativity and with spherical symmetry, it is 
possible to introduce the Kodama vector field $K^a$ 
\cite{Kodama}, which is used in the black hole literature as a 
substitute for timelike Killing vectors, which do not exist 
in time-dependent geometries. By contracting the Einstein 
tensor $G_{ab}$ 
with the Kodama vector one obtains an energy current $J^a 
\equiv G^{ab}K_b$ which, surprisingly, is covariantly 
conserved, $\nabla^b J_b=0$ \cite{Kodama} (a circumstance 
referred to as the ``Kodama miracle'' \cite{miracle}). As 
shown in Ref.~\cite{Haywardspherical}, the 
Misner-Sharp-Hernandez mass is the conserved Noether charge 
associated with the covariant conservation of the Kodama 
current $J^a$.

A third way of fixing the foliation involves using the  related Kodama gauge. 
The line element of a spherically 
symmetric spacetime can always be written as follows in a gauge 
 naturally using the areal radius $R$ (a geometrical quantity 
defined in a covariant way) as the radial coordinate:
\be\label{sphericallineelement}
ds^2=g_{00} (T, R) dT^2+g_{11}(T, R) dR^2 +R^2 
d\Omega_{(2)}^2 \equiv h_{ab}dx^a dx^b +R^2 d\Omega^2_{(2)} 
\,.
\ee
 (See Appendix \ref{appendix} for examples of known spacetimes 
expressed in the Kodama gauge.)
Eq.~(\ref{locusAH}) locating the apparent horizons becomes 
simply
\be
g^{RR}(T,R)=0 \,.
\ee
The Kodama vector is defined in a gauge-independent way as  
\cite{Kodama}
\be
K^a=\epsilon^{ab}\, \nabla_b R 
\ee
(where $\epsilon_{ab}$ is the volume form of the 2-metric 
$h_{ab}$) and it lies in the 
 2-space $(t,R)$ orthogonal to the 2-spheres of symmetry. 
Its components in coordinates $\left( T, R, \theta, 
\varphi \right)$ are
\be\label{Kodamavector}
K^{a} = \frac{1}{\sqrt{ |g_{TT}|g_{RR}}} \left( 
\frac{\partial}{\partial T} \right)^{a} \,.
\ee
$K^a$ identifies a preferred notion of time and becomes 
null on the apparent horizons, where $K_cK^c=0$, while it 
is timelike outside a black hole apparent horizon 
 and spacelike inside \cite{Kodama}. 
This behaviour is parallel to that of the timelike 
Killing field in stationary spacetimes with a Killing horizon. 
Ref.~\cite{miracle} advocates the gauge~(\ref{sphericallineelement}) 
on the basis of the simplicity it brings to the discussion of 
spherical apparent horizons. 
The gauge~(\ref{sphericallineelement}) using the areal 
radius $R$ as radial coordinate  seems 
motivated by the spherical symmetry and will be referred to 
as ``Kodama gauge''\footnote{This terminology is not standard.}
since the Kodama vector is naturally associated
with this gauge. We will now see that the 
characteristics of the Hawking mass in spherical symmetry 
support the choice of $R$ as the radial coordinate, and thus 
physically motivate the choice of the Kodama gauge.

\subsubsection{Hawking mass}

{\ There is a wide consensus on the relevance of the {\em dynamical} 
horizon (which is a spacelike apparent horizon \cite{Ashtekar:2003hk}) 
in non-stationary situations. For spherically symmetric dynamical  horizons,
the variation of energy constructed with the 
Ashtekar-Krishnan energy flux, introduced for dynamical 
horizons \cite{Ashtekar:2003hk} and computed for apparent 
horizons, is the variation of the Hawking/Misner-Sharp-Hernandez mass. 
The infinitesimal form of the area law is
\be 
\frac{dR}{2G} = dE_R\,, 
\ee
where we have restored Newton's constant $G$ for convenience.
$\kappa_R$ is the effective surface gravity 
of \cite{Ashtekar:2003hk}\footnote{The latter is different 
from the Hayward case, but they have the same stationary 
limit.} associated with the $R$-foliation as
\be 
\kappa_R \equiv \frac{1}{2R} 
\ee 
so that the 
infinitesimal form of the law is recast into the familiar form 
\be \label{1law2} 
\frac{\kappa_R \, dA}{8\pi G} = d E_R \,, 
\ee 
where $A$ is the area of a generic 
cross-section. Any other foliation leads to the law 
\be \label{1law3} 
\frac{\kappa_r \, 
dA}{8\pi G} = d E_{r} 
\ee 
provided that we define the effective surface gravity 
$\kappa_r$ of the $r$-foliation as
\begin{equation} 
\kappa_r = \frac{dr}{dR}\,\, 
\kappa_R \,, 
\ee 
and 
\be 
dE_r = \frac{dr}{dR} \, dE_R \,. 
\end{equation} 
One can choose 
different foliations for the apparent horizon in the spherically 
symmetric case, but the 
energy appearing in the area law reduces 
to the Hawking mass only if the areal radius $R$ 
is chosen as the radial coordinate \cite{Ashtekar:2003hk}. 
This gauge dependence can 
generically appear in other types of quasi-local masses such as the Brown-York mass 
\cite{Firouzjaee:2010ia} which are candidates to the role of mass for astrophysical 
systems \cite{Razbin:2012ve}. \\

Next, the surface ${\cal S}$ 
appearing in the definition of the Hawking mass is often 
identified with an apparent horizon (this is true, in 
particular, when identifying the Hawking mass $M_\text{H}$ 
with 
the internal energy of a black hole in the first law of 
horizon thermodynamics \cite{T1, T2, T3, T4}). 
Let us say that the foliation is 
now fixed by the Kodama gauge prescription given above. In 
general the 
unit normal to the apparent horizon ${\cal S}$ has 
components $n^{a}=\left( n^0, n^1, 0,0 \right)$ in the  
gauge~(\ref{sphericallineelement}), which are constrained 
by the normalization 
\be
g_{00}(n^0)^2+g_{11} (n^1)^2=-1
\ee  
with  $n^1 \neq 0$, so the surface ${\cal S}$ is not at 
rest in this gauge but is expanding or contracting 
radially. The Hawking mass is computed in the frame in 
which $u^{a} \propto K^{a}$, that is for the observer which 
has as time the Kodama time \cite{miracle}, 
and according to which the 
apparent horizon is not necessarily at rest. One can 
compute the mass $M_\text{H}$ on 2-spheres of symmetry 
outside the apparent horizon and then take the limit to 
this apparent horizon (in which, however, $K^a$ becomes  
a null vector \cite{Kodama}).

As shown by Hayward \cite{Haywardspherical}, $M_\text{MSH}$ 
is the conserved Noether charge associated with the 
conservation of the Kodama energy current $J^a$, so 
$M_\text{H}$ is the ``Newtonian'' mass in a frame in which 
there is no spatial flow of Kodama energy. This is not the 
frame in which ${\cal S}$ is at rest: in general the 
apparent horizon is not at rest but is in radial motion in the ``Kodama foliation'',
since $n^1 \neq 0$ in these 
coordinates.

In the Kodama gauge the Kodama current has components 
\be
J^{a}=\left( G^{00}K_0 , G^{10}K_0, 0, 0 \right) \,.
\ee
The component $J^1$ is proportional to $G^{TR}$ which, 
using the Einstein equations, is clearly proportional to a 
radial flow of ``energy'' $T^{TR}$.

 The Kodama foliation is therefore tightly linked to 
the widely used Misner-Sharp-Hernandez mass, which is one 
more argument in favour of its relevance, especially for 
thermodynamics considerations.

\subsubsection{Kodama gauge and horizon 
thermodynamics}
 
In discussions of the thermodynamics of apparent 
horizons, one way\footnote{In fact, this is the only method 
thus far  which is able to compute explicitly
 surface gravity and temperature of 
time-dependent apparent horizons in general 
(but spherical) spacetimes \cite{T1, T2, T3, T4}. 
Other methods have, thus 
far, produced results only for particular spacetime metrics 
\cite{Nielsenreview, other1, other2, other3}.} to 
compute the 
Hawking temperature consists of the so-called tunneling 
method, which uses the Kodama time as a preferred notion of 
time and the corresponding energy of a 
particle \cite{T1, T2, T3, T4, Helou1, Helou2}. An  
adiabatic approximation  expressing the fact that the 
apparent horizons
evolve slowly with respect to a ``background''  time scale 
of the dynamical spacetime is needed, but is not usually stated 
explicitly in the literature \cite{AlexJavad, mybook}. Clearly, 
fixing the foliation by making use of 
the areal radius $R$ and of the Kodama vector $K^a$ does not,
{\em per se}, prove that the Hawking temperature derived 
with the tunneling method is physical, nor that the first law of 
thermodynamics for apparent horizons based on this 
prescription is consistent and correct. The proof of these 
statements must come from somewhere else. The procedure 
presented here is only a physically plausible way of fixing 
the foliation for apparent horizons.

The tunneling method, used to derive the surface gravity and 
temperature of apparent horizons using the Kodama 
time, implicitly fixes the foliation and therefore selects 
apparent horizons. In other words, this method selects the 
family of observers $u^a_{(K)}$ associated with this 
foliation as those with 4-velocity proportional to the 
Kodama vector $K^a$ 
\cite{miracle} (see eqs.~(\ref{z1}) and~(\ref{z2}) below). 
However, the temperature of the black hole apparent 
horizon depends on the observer and this 
fact makes the entire thermodynamics of apparent horizons 
gauge-dependent, a property which, although intuitive {\em 
a posteriori}, is not usually noted even if there is debate 
on which is the ``correct'' temperature of apparent 
horizons.

\section{Relation between comoving and Kodama gauges}
\label{sec:4}
\setcounter{equation}{0}

Let us consider now the problem of finding the 
transformation from the comoving gauge~(\ref{comovinggauge}) 
to the Kodama gauge (\ref{sphericallineelement}). 
As shown below, the 
transformation cannot be written  in a completely explicit 
way. In order to find  the coordinate 
transformation, rewrite the line 
element~(\ref{comovinggauge}) in terms of the areal radius 
$R(t,r)$, using the fact that
\be\label{dr}
dr=\frac{dR -\dot{R}dt}{R'} \,.
\ee
Substituting eq.~(\ref{dr}) into the 
line element~(\ref{comovinggauge}) 
yields
\be
ds^2=-\left( \mbox{e}^{2\phi} -\frac{\dot{R}^2}{R'^2} \, 
\mbox{e}^{\lambda} \right) dt^2 +\frac{ 
\mbox{e}^{\lambda}}{R'^2} \, dR^2 -\frac{2\dot{R} 
\mbox{e}^{\lambda}}{R'^2} \, dtdR +R^2 d\Omega_{(2)}^2 \,.
\ee
The $dtdR$ cross-term can be eliminated by introducing a 
new time coordinate $T(t,R)$ such that
\be \label{KodamaTime}
dT = \frac{1}{F} \left( dt+\beta dR \right) \,,
\ee
where $F(t,R)$ is an integrating factor guaranteeing that 
$dT$ is an exact differential and $\beta( t, R)$ is a 
function to be determined. By substituting into the line 
element one obtains
\begin{eqnarray} 
ds^2 &=&-\left( \mbox{e}^{2\phi} -\frac{\dot{R}^2}{R'^2} \,
\mbox{e}^{\lambda} \right) F^2 dT^2 +2F \left[ \beta 
\left( \mbox{e}^{2\phi} -\frac{\dot{R}^2}{R'^2} \,
\mbox{e}^{\lambda} \right) -\frac{\dot{R}}{R'^2} \,  
\mbox{e}^{\lambda}  \right] dTdR \nonumber\\
&&\nonumber\\
&\, & + 
\left[ \frac{ \mbox{e}^{\lambda} }{R'^2} -\beta^2 \left( 
\mbox{e}^{2\phi} -\frac{\dot{R}^2}{R'^2} \,
\mbox{e}^{\lambda} \right) +2 \frac{\dot{R}}{R'^2} \,
\mbox{e}^{\lambda} \beta \right] dR^2 +R^2 d\Omega_{(2)}^2 
\,. \label{AA}
\end{eqnarray}
By setting 
\be\label{beta}
\beta(t,R)= \frac{ \dot{R} \, \mbox{e}^{\lambda}}{R'^2 
\left( \mbox{e}^{2\phi} - \mbox{e}^{\lambda}\dot{R}^2/R'^2 \right)} \,, 
\ee
the line element~(\ref{AA}) is diagonalized to the Kodama 
gauge
\begin{eqnarray}
ds^2 &=& - \left( \mbox{e}^{2\phi} -\frac{\dot{R}^2}{R'^2} 
\, \mbox{e}^{\lambda} \right) F^2 dT^2 + \frac{ 
\mbox{e}^{\lambda+2\phi} }{ R'^2 \mbox{e}^{2\phi} 
-\dot{R}^2 \,\mbox{e}^{\lambda} } \, 
dR^2 + R^2 d\Omega_{(2)}^2 \nonumber\\
&&\nonumber\\
&\equiv & g_{TT}(T, R) dT^2 +g_{RR}(T,R) dR^2 +R^2 
d\Omega_{(2)}^2 \,. \label{Kodamalineelement}
\end{eqnarray}
The metric components $g_{TT}$ and $g_{RR}$ are 
given implicitly as functions of $T$ and $R$. The 
integrating factor $F$ (which in general is not unique) 
must satisfy the equation
\be
\frac{\partial}{\partial R} \left( \frac{1}{F} \right) = 
\frac{\partial}{\partial t} \left( \frac{ \beta}{F} \right)  
\ee
which cannot be solved analytically except in trivial 
situations. 

In terms of the quantities $U$ and $\Gamma$ of 
eqs.~(\ref{U}) and (\ref{Gamma}), the Kodama line 
element~(\ref{Kodamalineelement}) reads
\be
ds^2= - \mbox{e}^{2\phi} \left( 1-\frac{U^2}{\Gamma^2} 
\right) F^2 dT^2 +\frac{ dR^2}{ \Gamma^2 -U^2} +R^2 
d\Omega_{(2)}^2 \,.
\ee   
The apparent horizons are located by the scalar equation 
$\nabla_c R \, \nabla^c R=0$ which in the Kodama gauge reduces to 
$g^{RR}=0$ and, of course, again gives $U=\pm \Gamma$. 
Therefore, we see explicitly that {\em the apparent 
horizons in the comoving gauge 
coincide with the apparent horizons in the Kodama gauge}. 
As already remarked, this conclusion is not surprising 
considering that the areal 
radius $R$ is a geometric quantity defined in an invariant 
way in spherical symmetry and the scalar equation 
$\nabla_cR \, \nabla^c R=0$ locating the apparent horizons is 
gauge-invariant.  
However, the 3-velocity of an apparent horizon with respect 
to the comoving observer ({\em i.e.}, to matter) is 
different from its 3-velocity with respect to the Kodama 
observers. The four-velocity of Kodama observers has 
components 
\be \label{z1}
u_{(K)}^{\mu} =\left( \frac{ 
\mbox{e}^{-\phi}}{F\sqrt{1-U^2/\Gamma^2} },0,0,0 \right)
\ee
in Kodama coordinates $\left( T,R, \theta, \varphi 
\right)$. Therefore, this vector is parallel to, but does 
not coincide with, the Kodama vector~(\ref{Kodamavector})
\be \label{z2}
K^{\mu} =\left( \frac{\Gamma \, \mbox{e}^{-\phi}}{F}, 0,0,0 
\right) \,.
\ee
The comoving observers have four-velocity given by
\be\label{u0c}
u_{(C)}^{\mu '}=
\frac{ \partial x^{\mu '}}{\partial x^{\mu}} \, u^{\mu}_{(C)} 
= \frac{ \partial x^{\mu '}}{\partial t} \, 
\mbox{e}^{-\phi} \,,
\ee
where $x^{\mu} \rightarrow x^{\mu '}$ is the coordinate 
transformation from comoving to Kodama coordinates and 
eq.~(\ref{ucomoving}) has been used. We have
\be
dT=\frac{1}{F} \left( dt+\beta dR\right)= \frac{\beta 
\dot{R}+1}{F} \, dt + \frac{\beta R'}{F}\, dr
\ee
and, using eq.~~(\ref{beta}), 
\be
\frac{\partial T}{\partial t} = \frac{1}{F}
\left[ 1+ \frac{ \dot{R}^2 \, \mbox{e}^{\lambda}}{R'^2 
\left( \mbox{e}^{2\phi} -\mbox{e}^{\lambda} \dot{R}^2/R'^2 
\right)} \right]=\frac{\Gamma^2}{F\left( 
\Gamma^2-U^2\right)} \,.
\ee
Eq.~(\ref{u0c}) and 
\be
u_{(C)}^{1'}=\frac{\partial R}{\partial t} \, 
\mbox{e}^{-\phi} \equiv U 
\ee
then give
\be
u_{(C)}^{\mu '}=\left( \frac{ \mbox{e}^{-\phi} }{F\left( 1-U^2/\Gamma^2 \right)} , U, 0,0 \right)
\ee
in Kodama coordinates. 

The scalar product between Kodama and comoving 4-velocities 
is 
\be
u_{(K)}^{a} u^{(C)}_a = -\frac{1}{ \sqrt{ 1-U^2/\Gamma^2}} 
=-\gamma ( v_\text{rel})\,,
\ee
where $v_\text{rel}$ is the instantaneous three-dimensional 
relative velocity between comoving and Kodama observers and 
$\gamma(  v_{rel})$ is the corresponding Lorentz factor. 
This (radial) 3-velocity has magnitude
\be
|v_\text{rel}| = \left|  \frac{U}{\Gamma} \right| 
\ee 
which depends on both time and radial location.  At the apparent horizon,
where \mbox{$U=-\Gamma$}, one has $\left|v_\text{rel}\right|=1$. 
The relative velocity of the Kodama observer with respect to the 
comoving observer is always equal to $\pm c$ at the horizon 
({\em i.e.} to $\pm 1$ since we are setting $c = 1$). This is 
logical since the Kodama vector is null at the apparent 
horizon.

Although 
the apparent horizons as determined by comoving and Kodama 
observers are the same, these surfaces will be perceived as 
different by these two families of observers because 
they are accelerated with respect to one another. If a 
temperature can be meaningfully assigned to time-dependent 
apparent horizons, the vacuum state of a field on the 
dynamical spacetime will be different with respect to 
Kodama and comoving observers. Moreover, the temperatures 
determined by these observers will be Doppler-shifted with 
respect to each other because the apparent horizon has  
different velocities relative to these observers. 

 In order to illustrate that, consider the velocity 
of the apparent horizon in the Kodama gauge. We have
\be
(v_\text{H}^{(K)})^2=\frac{g_{00}^2}{g_{11}^2} \, \left( \frac{dR}{dT}\right)^2=
\frac{ \mbox{e}^{-2\phi} \Gamma^2}{F^2 \left( \Gamma^2-U^2 \right)^2} 
\, \left( \frac{dR}{dT}\right)^2 \,.
\ee
Since
\begin{align}
\frac{dT}{dR} &= \frac{1}{F} \left( \frac{dt}{
\mbox{e}^{\phi} U dt+\mbox{e}^{\lambda/2} \Gamma dr} +
\beta \right)= \frac{1}{F} \left( \frac{\mbox{e}^{-\phi} }{U+\Gamma v} 
+\frac{ U \mbox{e}^{-\phi} }{\Gamma^2-U^2} \right) \nonumber \\
&=\frac{ \mbox{e}^{-\phi} }{F} \, \frac{ \Gamma  \left( \Gamma+Uv \right) }{ 
\left( U+\Gamma v \right)  \left(   \Gamma^2 -U^2\right)} \,,
\end{align}
we have at the horizon
\be
(v_\text{H}^{(K)})^2=
\frac{ \mbox{e}^{-2\phi} \Gamma^2}{F^2 \left( \Gamma^2-U^2 \right)^2} 
   \frac{F^2 \mbox{e}^{2\phi} \left( \Gamma^2-U^2\right)^2}{\Gamma^2}  
\left[  \frac{U(1-v)}{-U(1-v)} \right]^2 = 1 \,,
\ee
 which is expected since, again, the Kodama observer is null at the horizon. 
However the relative velocity $v_H^{(C)}$ of the horizon with respect to the comoving 
observer, given by eq.~\eqref{eq_vH_BH}, is in general 
not unity. The horizon indeed has a different velocity for the two observers.

The surface gravities $\kappa_{(K)}$ and $\kappa_{(C)}$ of 
Kodama and comoving observers will also differ and the 
horizon temperatures are given by 
$T_{(i)}=\kappa_{(i)}/(2\pi)$ (in geometrized units).\footnote{We 
can attribute to the surface gravity of an observer a 
``geometric temperature'' which can appear in a 
thermodynamic law. Hence the geometric temperatures of 
these observers are different. However we note that the Kodama vector 
temperature has two important properties: 1)~it reduces to  
the Killing  temperature for the case of an event 
horizon, and 2)~it is a key quantity in the tunneling method.}
Therefore, even though the previous considerations alleviate in practice 
the foliation-dependence problem of apparent horizons, the 
issue of associating physically meaningful temperatures 
and thermodynamics with these horizons is not addressed by 
our considerations above,  and it does not seem 
possible to alleviate the gauge-dependence of the thermodynamics 
of apparent horizons.

\section{Conclusions}
\label{sec:5}
\setcounter{equation}{0}

Nowadays apparent and trapping horizons are the common 
choice to describe black hole boundaries in numerical 
simulations of gravitational collapse and in works 
computing the waveforms of gravitational waves emitted 
during dynamical events in order to build banks of 
templates for the interferometric detection of 
gravitational waves. The recent {\em LIGO} observations of 
gravitational waves from binary black hole mergers 
\cite{LIGO, LIGO2} rely heavily on such waveforms and, 
therefore, on apparent horizons. Apparent and trapping 
horizons are much easier to locate than the teleological 
event horizons.  Therefore, it is hard to overemphasize the 
importance of apparent horizons in gravitational physics 
\cite{mybook}. However, apparent horizons depend on the 
spacetime slicing, as exemplified by the fact that one can 
even find foliations of the Schwarzschild spacetime which 
do not contain apparent horizons \cite{WaldIyer, 
SchnetterKrishnan}. The foliation-dependence of apparent 
horizons is a serious problem since in dynamical situations 
the apparent horizon takes on the role of the black hole boundary. 
The very existence of a black hole in dynamical 
spacetimes is therefore questioned if the existence of 
apparent horizons depends on the foliation, and realistic 
black holes are indeed dynamical due to the interaction 
with their environment and, ultimately, also because of 
Hawking radiation.

It seems that, at least in the presence of spherical 
symmetry as used in many studies of gravitational collapse, there 
should be a reasonable way out of this 
foliation-dependence problem. We restrict to this symmetry. 
Furthermore, it is natural to adopt a spherical foliation. 
Then, an apparent horizon is located by the scalar 
equation $\nabla_c R \, \nabla^c R=0$, where the areal radius 
$R$ is a gauge-independent quantity, which makes the 
(spherical) apparent horizons independent of the 
(spherically symmetric) foliation. This fact is reassuring since 
all observers associated with spherical foliations 
will agree on the 
existence of a black hole (identified with the region 
inside the same apparent horizon).

Both the comoving gauge (used mostly in numerical 
simulations) and the Kodama gauge (used mostly in the 
thermodynamics of apparent horizons and in Hawking 
radiation studies) are obvious candidates for a spherical 
foliation. Our discussion about the relation between the 
comoving and Kodama foliations will hopefully facilitate 
communication between the different communities working in 
these areas. We have checked explicitly that the apparent 
horizon is the same in these two gauges. However, due to 
their relative radial motion, comoving and Kodama observers 
perceive the same apparent horizon differently. While our 
discussion sheds some light on the foliation-dependence in 
the important case of spherical symmetry, the general 
situation of dynamical non-spherical geometries remains an 
open problem.

\section*{Acknowledgments}

The authors chiefly thank John Miller for useful discussions and comments on the manuscript.
V.F. thanks the University of Oxford, where this work was begun, 
for hospitality and the Natural Science and Engineering 
Research Council of Canada for financial support. The 
research leading to these results has received funding from the European 
Research Council under the European Community's Seventh Framework Program 
(FP7/2007-2013) StG-EDECS (Grant Agreement No. 279954) and from the ERC 
Advanced Grant 339169 ``Self-Completion''.

\newpage
\begin{appendices}
\appendix
\numberwithin{equation}{section}

\section{Appendix: Examples of comoving/Kodama gauges}
 \label{appendix}
\subsubsection*{Schwarzschild black hole}

As an example consider the Schwarzschild metric
\be
ds^2=-\left( 1-\frac{2M}{R}\right) dT^2+ 
\left( 1-\frac{2M}{R}\right)^{-1}dR^2 +R^2 d\Omega_{(2)}^2\,,
\ee
which, under this usual form, is naturally expressed 
in the Kodama gauge. Therefore, an observer at rest 
in the usual Schwarzschild coordinates is a Kodama observer, 
\emph{i.e.} an observer who remains on a given sphere of 
symmetry of areal radius $R$ (what we usually call an 
``accelerated observer'').
But the line element can be rewritten in the 
comoving gauge, using Lema\^itre coordinates, as
\be
ds^2=-dt^2+\frac{dr^2}{\left[ 3(r-t)/(4M) 
\right]^{2/3}} +(2M)^{2/3} \left[ \frac{3}{2} (r-t) 
\right]^{4/3} d\Omega_{(2)}^2
\ee
with $R=(2M)^{1/3}\left[ \frac{3}{2}(r-t) \right]^{2/3}$. 
An observer at rest in these coordinates is in free-fall 
towards the black hole.
Using the definition of the apparent horizon, it can be shown 
that the latter is located at $R=2M$ in both frames and that
we have \mbox{$M_\text{H}=M_\text{MSH}=M$}. 

\subsubsection*{Friedmann-Lema\^itre-Robertson-Walker cosmology}
The usual form of the FLRW line element
\begin{equation}
 ds^2=-dt^2 +\frac{a^2(t)}{1-kr^2} \, dr^2 + a^2(t)r^2 
d\Omega_{(2)}^2 \ ,
\end{equation}
with $a(t)$ the scale factor and $k$ the spatial curvature, 
is naturally written in comoving coordinates. By defining 
a new radial coordinate $R=a(t)r$, one obtains the so-called 
Pseudo Painlev\'{e}-Gullstrand\footnote{The name comes from 
the similarity of this metric with the Painlev\'e-Gullstrand 
form of the Schwarzschild metric. However, in the latter the 
$dR^2$ term appears with a unit coefficient, which is not 
the case here, whence the term ``pseudo''. See \cite{mybook} 
for the true Painlev\'e-Gullstrand form of the FLRW spacetimes.} 
line-element
\begin{equation}
ds^2 = -\left( 1-\frac{H^2R^2}{1-kR^2/a^2} \right) dt^2 -\frac{2HR}{1-kR^2/a^2}dtdR +\frac{dR^2}{1-kR^2/a^2} + R^2d\Omega_{(2)}^2 \ ,
\label{eq_Painleve_metric}
\end{equation}
where $H=\dot{a}/a$ is the Hubble parameter. 
However, although we have chosen the areal radius as 
our new radial coordinate, we have not yet obtained 
the Kodama gauge of eq.~\eqref{sphericallineelement}. 
That is because integral curves of the vector $\partial_t$ 
and of the Kodama  vector do not coincide \cite{miracle} 
(we do not have $K \propto \partial_t$).
One can get rid of the cross term by introducing a new time 
coordinate as in eq.~\eqref{KodamaTime},
\begin{equation}
ds^2 = -\left( 1-\frac{H^2R^2}{1-kR^2/a^2}  \right) F^2dT^2 
+\frac{dR^2}{1-kR^2/a^2-H^2R^2} + R^2d\Omega_{(2)}^2 \ ,
\label{eq_Painleve_metric2}
\end{equation}
with $a$, $H$ and $F$ implicit functions of $T$ and $R$ (see Ref.~\cite{mybook}). 
Now we indeed have that $K \propto \partial_T$, and we can refer 
to $T$ as the ``Kodama time'' of \cite{miracle}. An observer at rest 
in these coordinates is a Kodama observer, who remains on a given sphere of symmetry.
The very common line elements presented in this appendix are widely 
used and often preferred for their physical relevance. 
For the same reason, the apparent horizons uniquely defined in these gauges 
should be given a preferred status with respect to apparent horizons defined 
in non-symmetric gauges. 

\end{appendices}


\newpage
{\small \begin{thebibliography}{99}

\bibitem{Wald} R.M. Wald,  {\em General Relativity}  
(Chicago University Press, Chicago, 1984).

\bibitem{FrolovNovikov} V.P. Frolov and I.D. Novikov, {\em 
Black Hole Physics, Basic Concepts and New Developments} 
(Kluwer Academic, Dordrecht, 1998).

\bibitem{Poisson} E. Poisson, {\em A Relativist's Toolkit: 
The Mathematics of Black-Hole Mechanics} (Cambridge 
University Press, Cambridge, 2004).

\bibitem{Boothreview} I. Booth, ``Black hole boundaries'', 
{\em Can. J. Phys.} {\bf 83}, 1073 (2005).

\bibitem{Booth:2005ng} I.~Booth, L.~Brits, J.~A.~Gonzalez 
and C.~Van Den Broeck, ``Marginally trapped tubes and 
dynamical horizons'', {\em Class. Quantum Grav.} {\bf 23}, 
413 (2006).
  
\bibitem{Nielsenreview} A.B. Nielsen, ``Black holes and 
black hole thermodynamics without event horizons'', {\em 
Gen. Rel. Gravit.} {\bf 41}, 1539 (2009).

\bibitem{mybook} V. Faraoni, {\em Cosmological and Black 
Hole Apparent Horizons} (Springer, New York, 2015).

\bibitem{Diener2003} P. Diener, A new general purpose 
event horizon finder for 3D numerical spacetimes'', {\em 
Class. Quantum Grav.} {\bf 20}, 4901 (2003).

\bibitem{CohenPfeifferScheel} M.I. Cohen, H.P. 
Pfeiffer, and M.A. Scheel, ``Revisiting event horizon 
finders'', {\em Class. Quantum Grav.} {\bf 26}, 035005
(2009).

\bibitem{Thornburg} J. Thornburg, ``Event and apparent 
horizon finders for 3+1 numerical relativity'', {\em Living 
Rev. Rel.} {\bf 10}, 3 (2007). 
 
\bibitem{Lousto} M. Ponce, C. Lousto, and Y. Zlochower,  
``Seeking for toroidal event horizons from initially 
stationary BH configurations'', {\em Class. Quantum Grav.} 
{\bf 28}, 145027 (2011).

\bibitem{BaumgarteShapiro} T.W. Baumgarte and S.L. Shapiro, 
``Numerical relativity and compact binaries'', {\em Phys. 
Rept.} {\bf 376}, 41 (2003).

\bibitem{Chuetal} T. Chu, H.P. Pfeiffer, and M.I. Cohen, 
``Horizon dynamics of distorted rotating black holes'', 
{\em Phys. Rev. D} {\bf 83}, 104018 (2011).

\bibitem{LIGO} B.P. Abbott {\em et al.} (LIGO Scientific 
Collaboration and Virgo Collaboration), ``Observation of 
gravitational waves from a binary black hole merger'', {\em 
Phys. Rev. Lett.} {\bf 116}, 061102 (2016).

\bibitem{LIGO2} LIGO Scientific Collaboration and Virgo Collaboration,
``Binary black hole mergers in the first Advanced LIGO observing run'', 
arXiv:1606.04856.

\bibitem{Rindler} W. Rindler, ``Visual horizons in 
world-models'', {\em Mon. Not. R. Astr. Soc.} {\bf  116}, 
663 (1956) [reprinted in {\em Gen. Rel. Gravit.} {\bf 34}, 
133 (2002)].

\bibitem{WaldIyer} R.M.Wald and V. Iyer, ``Trapped surfaces 
in the Schwarzschild geometry and cosmic censorship'', {\em 
Phys. Rev. D} {\bf 44}, R3719 (1991).

\bibitem{SchnetterKrishnan} E. Schnetter and B. 
Krishnan, ``Non-symmetric trapped surfaces in the 
Schwarzschild and Vaidya spacetime'', {\em Phys. Rev. D} 
{\bf 73}, 021502 (2006).

\bibitem{Firouzjaee:2010ia} J.T. Firouzjaee, M.P. Mood, and 
R. Mansouri, ``Do we know the mass of a black hole? Mass of 
some cosmological black hole models,'' {\em Gen. Rel. 
Gravit.} {\bf 44}, 639 (2012).

\bibitem{Lemaitremass} V. Faraoni, ``Lema\^itre  model 
and cosmic mass'', {\em Gen. Rel. Gravit.} {\bf 47}, 
84 (2015).

\bibitem{MSH2} W.C. Hernandez and C.W. Misner, ``Observer 
time as a coordinate in relativistic spherical 
hydrodynamics'', {\em Astrophys. J.} {\bf 143}, 452 (1966).

\bibitem{Hawking} S. Hawking, ``Gravitational radiation 
in an expanding universe'', {\em J. Math. Phys.} {\bf 9}, 
598 (1968).

 
\bibitem{MSH} C.W. Misner and D.H. Sharp,    
``Relativistic equations for adiabatic, spherically 
symmetric gravitational collapse'',  {\em Phys. Rev.} {\bf 
136}, B571 (1964). 

\bibitem{Visser}
  M.~Visser,
  ``Physical observability of horizons,''
 {\em Phys. Rev. D} {\bf 90}, 127502 (2014).

\bibitem{Hayward:1993wb}   
  S.~A.~Hayward,
  ``General laws of black hole dynamics,''
  {\em Phys. Rev. D} {\bf 49}, 6467 (1994).

\bibitem{Hayward:1997}      
  S. Hayward and M. Kriele, 
  ``Outer trapped surfaces and their apparent horizon,'' 
  {\em J. Math. Phys.} {\bf 38}, 1593 (1997).

\bibitem{Eardley:1997hk}    
  D.~M.~Eardley,
  ``Black hole boundary conditions and coordinate conditions,''
  {\em Phys. Rev. D} {\bf 57}, 2299 (1998).
  
\bibitem{BenDov:2006vw}     
  I.~Ben-Dov,
  ``Outer Trapped Surfaces in Vaidya Spacetimes,''
  {\em Phys. Rev. D} {\bf 75}, 064007 (2007).

\bibitem{Ashtekar:2005ez} 
  A.~Ashtekar and G.~J.~Galloway,
  ``Some uniqueness results for dynamical horizons,''
  {\em Adv. Theor. Math. Phys.} {\bf 9} no.1, 1 (2005).

\bibitem{Booth:2005qc}  
  I.~Booth,
  ``Black hole boundaries,''
  {\em Can. J. Phys.} {\bf 83}, 1073 (2005).
      
\bibitem{Bengtsson:2010tj} 
  I.~Bengtsson and J.~M.~M.~Senovilla,
  ``The Region with trapped surfaces in spherical symmetry, its core, and their boundaries,''
  {\em Phys. Rev. D} {\bf 83}, 044012 (2011) .
  
\bibitem{Szabados} L.B. Szabados, ``Quasi-local 
energy-momentum and angular momentum in general 
relativity'', {\em Living Rev. Rel.} {\bf 12}, 
4 (2009).

\bibitem{Haywardspherical} S.A. Hayward, ``Gravitational 
energy in spherical symmetry'', {\em Phys. Rev. D} {\bf 53}, 
1938 (1996).

\bibitem{HelouMuscoMiller2016} A. Helou, I. Musco, and J.C.  
Miller, ``Causal nature and dynamics of trapping horizons 
in black hole collapse and cosmology'', arXiv:1601.05109.

\bibitem{Kodama} H. Kodama, ``Conserved energy flux 
from the spherically symmetric system and the back reaction 
problem in the black hole evaporation'', {\em Progr. Theor. 
Phys.} {\bf 63}, 1217 (1980).  

\bibitem{miracle} G. Abreu and M. Visser, ``Kodama time: 
geometrically preferred foliations of spherically symmetric 
spacetimes'', {\em Phys. Rev. D} {\bf 82}, 044027 (2010).

\bibitem{Ashtekar:2003hk} A. Ashtekar and B.~Krishnan, 
``Dynamical horizons and their properties,'', {\em Phys. 
Rev. D} {\bf 68}, 104030 (2003).

\bibitem{Razbin:2012ve} 
M.~Razbin, J.T.~Firouzjaee, and R.~Mansouri,
``Relativistic rotation curve for cosmological structures'', 
{\em Int.\ J.\ Mod.\ Phys.\ D} {\bf 23},  1450074 (2014).

\bibitem{T1} M. Angheben, M. Nadalini, L. Vanzo, and S. 
Zerbini, ``Hawking radiation as tunneling for extremal and 
rotating black holes'', {\em J. High Energy Phys.} {\bf 
05}, 014 (2005).

\bibitem{T2} R. Di Criscienzo, M. Nadalini, L. Vanzo, S. 
Zerbini, and G. Zoccatelli, ``On the Hawking radiation as 
tunneling for a class of dynamical black holes, {\em Phys. 
Lett. B} {\bf 657}, 107 (2007).

\bibitem{T3} R. Di Criscienzo, S.A. Hayward, M. Nadalini, L. Vanzo, and 
S. Zerbini, ``Hamilton-Jacobi method for dynamical horizons in different 
coordinate gauges'', {\em Class. Quantum Grav.} {\bf 27}, 015006 (2010).

\bibitem{T4} L. Vanzo, G. Acquaviva, and R. Di Criscienzo, 
``Tunnelling methods and Hawking's radiation: achievements 
and prospects'', {\em Class. Quantum Grav.} {\bf 28}, 
183001 (2011).
 
\bibitem{other1} H. Saida, T. Harada, and H. Maeda, ``Black hole evaporation 
in an expanding universe'', {\em Class. Quantum Grav.} {\bf 24}, 4711 
(2007).

\bibitem{other2} B.R. Majhi, ``Thermodynamics of Sultana-Dyer black 
hole'', {\em J. Cosmol. Astropart. Phys.} {\bf 1405}, 014 (2014).

\bibitem{other3} K. Bhattacharya and B.R. Majhi, 
``Temperature and thermodynamic structure of Einstein's 
equations for a cosmological black hole'', {\em Phys. Rev. 
D} {\bf 94}, 024033 (2016).

\bibitem{Helou1} P.~Bin\'etruy and A.~Helou, ``The Apparent 
Universe,''  {\em Class. Quantum Grav.}  {\bf 32}, 205006 
(2015).

\bibitem{Helou2} A.~Helou, ``Dynamics of the Cosmological 
Apparent Horizon: Surface Gravity \& Temperature'', 
arXiv:1502.04235 [gr-qc].
  
\bibitem{AlexJavad} A.B. Nielsen and J.T. Firouzjaee, 
``Conformally rescaled spacetimes and Hawking radiation'', 
{\em Gen. Rel. Gravit.} {\bf 45}, 1815 (2013).


\end{thebibliography}}
\end{document}